\documentclass[12pt]{article}
\begin{document}

\begin{center}
{\large Interference Between a Large Number of Independent Bose-Einstein 
Condensates}

\

S. Ashhab

{\footnotesize Department of Physics, Ohio State University, 174 West 
Eighteenth Avenue, Columbus, Ohio 43210}

\vspace{2cm}

{\bf Abstract}
\end{center}

We study theoretically the interference patterns produced by the overlap 
of an array of Bose-Einstein condensates that have no phase coherence 
among them. We show that density-density correlations at different 
quasimomenta, which play an important role in two-condensate interference, 
become negligible for large $N$, where $N$ is the number of overlapping 
condensates. In order to understand the physics of this phenomenon, it is 
sufficient to consider the periodicity of the lattice and the statistical 
probability distribution of a random-walk problem. The average visibility 
of such interference patterns decreases as $N^{-1/2}$ for large $N$.

\newpage

Since the first experimental realizations of Bose-Einstein condensates 
(BEC) in the alkali-atomic gases \cite{Anderson1,Davis,Bradley}, one of 
the first goals was to demonstrate the coherence of the matter waves in a 
BEC. That goal was first achieved in the classic interference experiment 
by Andrews {\it et. al.} \cite{Andrews}. In the following years, 
interference effects proved to be a valuable experimental tool in the 
study of cold atomic gases \cite{Anderson2,Bloch,Orzel,Greiner1,Greiner2}. 
For example, they were used to demonstrate phase coherence within a single 
condensate \cite{Bloch}, study squeezed states of a BEC in a 
double-well potential \cite{Orzel}, identify the phase transition 
between the superfluid and Mott-insulator phases of a Bose gas in an 
optical lattice \cite{Greiner1} and observe the dynamics of matter wave 
fields \cite{Anderson2,Greiner2}. With the increased interest in atomic 
gases trapped in optical lattices, interference experiments will 
undoubtedly provide a useful tool to probe the coherence properties of 
these systems. It is therefore important to understand the physics of 
interference patterns produced following the release and expansion of such 
atomic gases.

The phenomenon of interference between BEC's manifests itself as the 
observation of a periodic modulation of the density caused by the 
overlap of two or more condensates. When two condensates with 
an equal number of atoms are made to overlap, such a density modulation 
occurs in every run of the experiment with 100\% visibility, i.e. the 
density is ideally the square of a sinusoidal function, up to an envelope 
function resulting from the spatial finiteness of the system 
\cite{Andrews}. Therefore, one can say that there exist strong 
correlations that dictate that the density must vanish at the midpoint 
between two successive maxima. If one starts with two independent 
condensates, i.e. if the relative phase between the two condensates is 
initially unknown, the location of interference maxima takes a random 
value, keeping the distance between successive maxima fixed by the length 
and time scales in the experimental setup \cite{Andrews,Javanainen}. Note 
that in the context of interference experiments, the term ``independent 
condensates'' describes both a statistical mixture of states with 
well-defined but unknown relative phases and Fock states, where one starts 
with well-defined values of the atom numbers in each condensate. In the 
latter case, the location of the emerging interference maxima takes a 
random value, just like any quantum variable whose wave function collapses 
at the time of measurement.

If three condensates are made to overlap, one must consider two relative 
phases, which can be taken as the relative phase between condensates 1 
and 2 and that between condensates 2 and 3. It is therefore intuitive to 
think that the overlap of three condensates will still produce 
interference fringes, but with reduced visibility due to the randomness of 
the two degrees of freedom. Similarly, one would expect the interference 
between a larger number of condensates to produce interference fringes 
with a value of visibility that decreases and eventually vanishes with 
increasing number of condensates. In a recent experiment, Hadzibabic {\it 
et. al.} studied the interference of about 30 condensates, and they 
observed density modulations with 34\% average visibility 
\cite{Hadzibabic}. At first sight, that result might seem to contradict 
the above-presented intuitive guess. We shall demonstrate that there is no 
contradiction between the experiment and our intuition. Our intuition 
might fail, however, in predicting how quickly the visibility decreases 
with increasing number of condensates.

We shall consider the experimental situation studied in Ref. 
\cite{Hadzibabic}. In that experiment, a BEC is loaded into a 
one-dimensional optical lattice such that a large number of lattice sites 
are filled with atoms (Note that the lattice sites are not 
equally-populated, but each one contains a large number of atoms 
\cite{NumberOfAtoms}). With the choice of parameters in their setup, apart 
from determining the occupation of the different lattice sites, 
interatomic interactions can be neglected throughout the experiment. 
Furthermore the height of the optical lattice potential is substantially 
larger than the kinetic energy of an atom in the ground state of a given 
lattice site. Therefore, we find that the (single-particle) wave function 
describing the condensate is well-approximated by a sum of Gaussians 
centred at the different potential minima. Note that all the Gaussians 
have the same width, namely that given by the harmonic oscillator length 
near the potential minima. The trapping potential is then turned off, and 
the cloud expands ballistically. After a long expansion time, a picture is 
taken of the density distribution. In Ref. \cite{Hadzibabic} the 
experiment was analysed theoretically by taking the initial wave function 
and propagating it in time to obtain numerically the density distribution 
after the expansion. In this Paper we shall try to identify the origin of 
the different physical aspects of the experiment analytically, and we 
verify that identification with some numerical calculations. Unlike the 
analysis of Ref. \cite{Hadzibabic}, in order to avoid complications that 
do not have any qualitative effect on the behaviour of main interest to 
us, we shall not worry about the finite resolution of the imaging device. 
Such effects can be taken into account relatively easily.

One can gain a good amount of insight into the physics of the problem by 
considering the different relevant length scales and what they correspond 
to before and after the expansion of the atomic cloud (here we assume that 
the expansion lasts long enough so that any structure seen in the final 
density distribution is larger than the original size of the cloud before 
the expansion). The relation between each length scale $x$ before the 
expansion and the length scale it produces after the expansion $X$ is 
given by $X\sim\hbar t/mx$. The size of the wave function inside 
each well in the optical lattice $\xi$ gives the overall size of the cloud 
after the expansion $\Xi\sim\hbar t/m\xi$. That latter length scale gives 
the Gaussian envelope of the density distribution after the expansion. The 
distance between lattice sites $d$ gives the period of of the density 
modulation inside the Gaussian envelope $D\sim\hbar t/md$. The size of the 
entire cloud before the expansion $l$ gives the length scale of structure 
in the periodic function $L\sim\hbar t/ml$, i.e. changes in the density 
after the expansion can only occur on length scales of the order of or 
larger than $L$. In the following paragraphs we shall revisit the above 
considerations more quantitatively and develop a simple framework for 
thinking about the problem at hand.

First, we look at the wave function before the expansion. As explained 
above, it can be approximated by:

\noindent
\begin{equation}
\psi (x) = \sum_{j} \alpha_j e^{i\theta_j} e^{-(x-jd)^2/2\xi^2},
\end{equation}

\noindent
where $\alpha_j e^{i\theta_j}$ are the amplitudes of the wave function at 
the different lattice sites, labelled by $j$ ($j\!\! =\!\! 1,2,...,N$, 
where $N$ is the number of overlapping condensates). The Fourier transform 
of that wave function can be calculated straightforwardly to give the wave 
function in momentum space:

\noindent
\begin{eqnarray}
\Phi (p) & = & \frac{1}{\sqrt{2\pi\hbar}} \int_{-\infty}^{\infty} 
e^{-ipx/\hbar} \psi(x) {\rm d}x
\nonumber
\\
& = & \frac{\xi}{\sqrt{\hbar}} \left( \sum_{j=1}^{N} \alpha_j 
e^{i\theta_j} e^{-ipdj/\hbar} \right) e^{-p^2\xi^2/2 \hbar^2}
\label{eq:Fourier_Transform}
\\
\left|\Phi(p)\right|^2 & = & \frac{\xi^2}{\hbar} \left( \sum_{jj'=1}^{N} 
\alpha_j \alpha_{j'} e^{i\theta_j-\theta_{j'}} e^{-ipd(j-j')/\hbar} 
\right) e^{-p^2\xi^2/\hbar^2}
\nonumber
\\
& = & \frac{\xi^2}{\hbar} \left( \sum_{n=0}^{N-1} A_n \cos 
(\frac{ndp}{\hbar}+\delta_n) \right) e^{-p^2\xi^2/\hbar^2},
\label{eq:Density}
\end{eqnarray}

\noindent where

\noindent
\begin{eqnarray}
A_o & = & \sum_{m=1}^{N} |\alpha_m|^2 \: , \; \delta_o=0
\nonumber
\\
A_n & = & 2 \left| \sum_{m=1}^{N-n} \alpha_m \alpha_{m+n} e^{i 
(\theta_m-\theta_{m+n})} \right| \hspace{1cm} , \, n=1,2,...,N-1
\label{eq:A_n}
\end{eqnarray}

\noindent
and $\delta_n$ is the argument (i.e. phase angle) of the expression inside 
the absolute value brackets in the expression for $A_n$. Since the density 
distribution in real space after a long time of ballistic expansion 
reflects the density distribution in momentum space before the expansion, 
the above expression must contain all the information about the produced 
interference patterns. As explained in the previous paragraph, we can 
immediately see that the last exponential factor in Eq. 
(\ref{eq:Fourier_Transform}) provides the Gaussian envelope of the density 
distribution. The term inside the brackets in Eq. 
(\ref{eq:Fourier_Transform}), to which we shall refer as $\phi(p)$, 
describes the periodic function inside the Gaussian envelope [note that 
$\phi(p+2\pi\hbar/d)=\phi(p)$]. Since the largest value of $m$ is $N-1$, 
$p$ must change by at least $\sim\hbar/(Nd)$ in order to see any 
substantial change in the density distribution [note that $Nd$ is the size 
of the entire cloud before the expansion]. It is perhaps easiest to think 
about the above analysis in term of Bloch states, which describe the wave 
function of a particle in a periodic potential \cite{Ashcroft}. One can 
then think of $\phi(p)$ as describing the probability amplitude of a given 
atom to be in the quasimomentum state $p$ in the lowest Bloch band. It now 
suffices to consider the discrete set of momenta $p_j=2\pi\hbar j/(Nd)$, 
where $j$ runs over the integers from $-N/2+1$ to $N/2$. The scale for 
density variations now becomes identified as the separation between 
adjacent quasimomentum states. In other words, once we have calculated 
$\phi(p)$ at the discrete set of quasimomenta, we can connect the points 
with a smooth curve to find $\phi(p)$ for any value of $p$.

In order to demonstrate the unique features of the two-condensate 
interference problem, we discuss it in some more detail (we take $N=2$ and 
$\alpha_1=\alpha_2\equiv\alpha$). The language of Bloch states is 
not suited to describe such a small value of $N$. However, if one takes 
any two points in momentum space that are separated by $\pi\hbar/d$, which 
corresponds to the distance between quasimomentum states for $N=2$, one 
finds a clear density-density correlation, which can be expressed as:

\noindent
\begin{equation}
\frac{\langle \left( |\phi(p)|^2-\bar{\rho} \right) 
\left( |\phi(p+\pi\hbar/d)|^2 -\bar{\rho} \right) \rangle}{\bar{\rho}^2} = 
-\frac{1}{2},
\end{equation}

\noindent 
where $\bar{\rho}$ is the average value of $|\phi(p)|^2$ and represents 
the average density of the interference pattern, neglecting the Gaussian 
envelope. The effect of that correlation is displayed in its most 
spectacular form when one starts with a Fock state \cite{Javanainen}. In 
that case, the detection of the first few atoms, combined with 
density-density correlations, is the mechanism responsible for the 
determination of the relative phase between the two condensates. Another 
manifestation of that density-density correlation is the vanishing density 
at the interference minima. The correlations can also be seen using the 
treatment of the previous paragraph. From Eq. (\ref{eq:A_n}) we find that 
$A_o=A_1= 2 |\alpha|^2$, which says that the visibility must be 100\% in 
every run of the experiment. If one identifies the points of maximum and 
minimum density as the preferred quasimomentum states, the densities are 
automatically set at $2\bar{\rho}$ and zero, repectively. This type of of 
density-density correlations at the different quasimomenta decreases if we 
increase $N$. However, more intricate correlations appear as $N$ is 
increased, corresponding to the increasing number of $A_n$'s. Therefore, 
one might think that it is important to take into account the effects of 
all those correlations to describe the interference patterns correctly. We 
shall show below that neglecting all such correlations still yields a good 
approximation of the produced interference fringes.

We now proceed with the numerical calculations and comment on the results 
as we move along. Fig. 1 shows the average visibility $V$ as a function of 
the number of overlapping condensates $N$. Each point in Fig. 1 represents 
the average value of $10^3$ runs of the simulation. First, we perform what 
we refer to as the exact calculation. In each run, we generate random 
values for the phases of the condensates in the different lattice sites. 
From those phases we can calculate the probability amplitude $f(p)$ for an 
atom to be in any of the $N$ quasimomentum states $p$ [$p\! =\! 2\pi\hbar 
j/(Nd)), \, j=-N/2+1,...,N/2$]. We then find a least-squares fit to the 
squares of those amplitudes (i.e. $|f(p)|^2$) of the form:

\noindent
\begin{equation}
h(p) = h_o \left\{ 1 + V \cos \left( \frac{pd}{\hbar} + \delta \right) 
\right\}.
\end{equation}

\noindent
We perform the exact calculation for both a Thomas-Fermi (TF) density 
distribution, where the initial occupation of the lattice sites has the 
form of an inverted parabola, and for the homogeneous case, where all the 
lattice sites are equally-populated. The reason why the average visibility 
is higher in the TF case than the uniform case for the same number of 
condensates is that in the TF case the edge condensates, which have a 
small number of atoms, have a small effect on the produced interference 
patterns, and that results in a smaller effective number of condensates. 
In the remainder of this Paper, we focus on the homogeneous case. When 
calculating the values $f(p)$ in the above simulations, we use the same 
values of the condensate phases for all the points $p$. We note, however, 
that the number of points that we generate in the function $f(p)$ is equal 
to the number of randomly-generated phases. Moreover, as the number of 
condensates $N$ increases, correlations in $f(p)$ at different points $p$ 
become more subtle. For example, two-point correlations of the form 
$\langle (|f(p)|^2 |f(p')|^2 \rangle / \langle |f(p)|^2 \rangle \langle 
|f(p')|^2 \rangle -1$ decrease and approach zero as $N\rightarrow\infty$. 
It is then natural to ask whether it is necessary to keep track of the 
condensate phases when we calculate $f(p)$ at the different points $p$. We 
answer that question by generating another set of simulations (for the 
homogeneous case), but now for each value of $p$ we calculate $f(p)$ by 
summing $N$ terms of the form $e^{i\theta}$ with randomly chosen values of 
the variable $\theta$. We can see from Fig. 1 that for 
$N\stackrel{>}{\sim}20$ the two sets become almost indistinguishable, 
apart from statistical fluctuations. Although this calculation does not 
correspond to any simple physical statistical ensemble in the present 
context, it has the advantage of simplifying our thinking about the 
problem. When visualising the produced interference patterns, we no longer 
have to keep track of how the phase of the matter-wave field from each 
condensate changes as we move across the (imaginary) detection screen. We 
can simply think of a function where to each of the $N$ different values 
of quasimomentum on the $x$-axis we assign a randomly-generated number as 
the value of the function (we shall use that idea to give a simple 
derivation of the large $N$ behaviour of the average visibility $V$ 
below). The problem can be simplified a little bit further as follows. For 
large $N$ the problem of adding $N$ terms of the form $e^{i \theta}$ with 
random values of $\theta$ is a two-dimensional random walk problem. As is 
well-known, the large-$N$ random walk problem in two dimensions has the 
probability density \cite{McCrea}:

\noindent
\begin{equation}
g(r) = \frac{\pi}{2R^2} r e^{-\pi r^2/4R^2},
\label{eq:Random_Walk_Dist}
\end{equation}

\noindent
where the variable $r$ is the distance from the origin, and $R$ is the 
mean value of $r$ [Note that $r$ corresponds to $|f(p)|$ in the present 
problem]. We generate a fourth set of points where we now generate each 
point $|f(p)|$ from the probability distribution in Eq. 
(\ref{eq:Random_Walk_Dist}). From Fig. 1, we can see that this new set of 
points agrees with the exact calculation just as well as the previous 
calculation, even for $N$ as low as 11, where approximating the 
probability distribution of the random-walk problem by Eq. 
(\ref{eq:Random_Walk_Dist}) is not expected to be very accurate. 
For further comparison, we plot in Fig. 2 the probability desnity $P(V)$ 
to find a certain value of the visibility $V$ for $N=11$ and $N=51$. Fig. 
2 shows the results for both the exact calculation and the calculation 
where correlations are neglected. We do that by running the simulation 
$10^5$ times, and then distributing the obtained values of $V$ into a 
histogram. Again, we find very good agreement between the two 
calculations. We note that if we look at higher-frequency components in 
the density distribution, i.e. terms corresponding to higher values of $n$ 
in Eq. (\ref{eq:Density}), we expect to see less agreement between our 
approximation and the exact calculation. The reason is that $\langle A_n 
\rangle$ decreases substantially as $n$ becomes of order $N$ in the exact 
calculation, whereas our approximation gives a value of $\langle A_n 
\rangle$ that is independent of $n$. The effect of those high frequency 
components, however, is rather difficult to see from a simple view of the 
interference patterns.

For completeness, we make the following two observations about the results
of our simulations: (1) Using an unrestricted least-squares fit, we
obtained values of the visibility greater than 1. That happened about 5\%
of the runs for $N=11$ (homogeneous case), and did not occur for $N>40$.
In those cases we have used the value 1. The difference in the average 
visibility between the corrected and uncorrected sets was always under 
2\%. (2) In all the data sets, the standard deviation in the visibility 
converged to around 0.52 of the mean for large $N$. That value can be 
obtained analytically for the exact calculation by making the observation 
that the statistics of the measured values of the visibility, as given by 
Eq. (\ref{eq:A_n}), is described by a two-dimensional random-walk problem. 
Using Eq. (\ref{eq:Random_Walk_Dist}), one can see that the ratio of the 
standard deviation to the mean of the distance in that problem is given by 
$\sqrt{4/\pi-1}=0.523$.

We now derive the large $N$ behaviour of $V$. One can do that by 
considering Eq. (\ref{eq:A_n}). It is straightforward to see that:

\noindent
\begin{equation}
\frac{\langle A_n^2 \rangle}{\langle A_o \rangle^2} = 4\frac{N-n}{N^2} 
\hspace{1cm} , \, n=1,2,...,N-1.
\end{equation}

\noindent
Moreover, since the calculation of $\langle A_n \rangle$ reduces to a 
two-dimensional random walk problem in the large $N$ limit [with $(N-n)\gg 
1$], we find that $\langle A_n \rangle^2/\langle A_n^2 \rangle= \pi/4$, 
which gives the average visibility as $V = \langle A_1 \rangle / 
\langle A_o \rangle \rightarrow \sqrt{\pi/N}$. We plot that function on 
Fig. 1 for comparison with our simulations. We now derive the $N^{-1/2}$ 
behaviour using the approach where we neglect multiple-point correlations 
in $f(p)$. As explained above, $f(p)$ is periodic in $p$ with period 
$2\pi\hbar/d$. We take that interval and divide it into $K$ regions of 
equal length, where $K\gg 1$. We now take $N$ to be a large multiple of 
$K$, i.e. $N=MK$ with $M\gg 1$, so that each region contains $M$ points 
where the function $f(p)$ is to be evaluated. For large $K$ and $M$, we 
can replace all the points in each region by one point that represents the 
average value $F$ of the function in that region. That coarse graining 
procedure adds a negligibly small correction to the value of the 
visibility that we obtain in the density fit. At each point, $f(p)$ is 
generated from a probability distribution with mean $\bar{f}$ and standard 
deviation $\Delta f$. Note that the ratio $\Delta f/\bar{f}$, which can be 
obtained from Eq. (\ref{eq:Random_Walk_Dist}), is independent of $N$. With 
a simple statistics argument, we find that $\Delta F/\bar{F} = 
(1/\sqrt{M}) \Delta f/\bar{f}$, where $\bar{F}$ and $\Delta F$ are the 
mean and standard deviation of $F$, respectively. The quantity $\Delta 
F/\bar{F}$ measures the relative deviations of the density from its mean. 
Since it is those deviations that produce the finite density modulation, 
the average visibility must also be proportional to $1/\sqrt{M}$ for large 
$M$. It therefore follows that $V$ falls off as $N^{-1/2}$ for large 
values of $N$.

We pause for a moment to comment on one of the calculations by Hadzibabic 
{\it et. al.} \cite{Hadzibabic}. They calculated numerically the minimum 
and maximum densities of every shot and averaged the results over many 
shots. They found that the average minimum density $\rho_{\rm min}\sim 
N^{-1}$ and the average maximum density $\rho_{\rm max}\sim\ln N$, so that 
$\rho_{\rm min}/\rho_{\rm max}\rightarrow 0$ as $N\rightarrow\infty$. In 
our approach we can obtain the large $N$ bahaviour of those quantities 
straightforwardly by requiring that:

\noindent
\begin{eqnarray}
N \int_{0}^{\sqrt{\rho_{\rm min}}} g(r) {\rm d}r \sim 1
\\
N \int_{\sqrt{\rho_{\rm max}}}^{\infty} g(r) {\rm d}r \sim 1.
\end{eqnarray}

\noindent
Solving those equations in the limit $N\rightarrow\infty$ gives the 
above asymptotic behaviour. In Ref. \cite{Hadzibabic} that result is 
described as being an increasing contrast with increasing $N$. However, 
one must be careful when using that definition of contrast, since the 
visibility of interference fringes decreases with increasing $N$, as we 
have shown above and as mentioned in Ref. \cite{Hadzibabic}. Note also 
that the width of those features, i.e. the density maxima and minima, 
decreases with increasing number of condensates.

We finally comment briefly on the three-dimensional case. As above, using 
arguments of Bloch states we can see that the density distribution after 
the expansion will have the same symmetries as the reciprocal lattice. For 
example, if the real lattice is cubic, the reciprocal lattice will also 
be cubic, and so will be the density distribution after the expansion. 
In the sense that an interference pattern is a periodic modulation of the 
density, we would then expect to see interference patterns in any number 
of dimensions. One complication arises when one tries to image such 
interference patterns. With present-day imaging techniques, one can only 
obtain two dimensional pictures. Therefore, if one simply shines a laser 
beam at the condensate and looks at the shadow of that beam, only the 
total density along the imaging direction is measured. That has the effect 
of reducing the visibility of the interference pattern from its value for 
a two-dimensional lattice. That can be seen by considering first the 
two-dimensional case, which is produced by generating a set of random 
numbers with mean value $\bar{f}$ and standard deviation $\Delta f$ at the 
quasimomentum states. In the three-dimensional case, we instead have to 
add $M$ such randomly-generated numbers at each quasimomentum state (in 
the reduced two-dimensional space; here we consider a cubic lattice for 
simplicity), where $M$ is the number of lattice sites in the third 
dimension. The mean of the sum is given by $M \bar{f}$, whereas the 
standard deviation of the sum is given by $\sqrt{M} \Delta f$. Therefore 
the deviations from the mean in the measured density will be reduced by a 
factor $M^{-1/2}$, leading to a reduced visibility. It is possible, 
however, to image slices of the expanded cloud as explained in Ref. 
\cite{Andrews}. Using that technique, the three dimensional density 
distribution should be accessible.

In conclusion, we have considered the interference patterns produced by 
the overlap of an array of independent Bose-Einstein condensates. We have 
shown that density-density correlations at different quasimomenta do not 
play any significant role in producing the interference fringes, in 
contrast to what might be intuitively expected. In order to understand the 
main features of the produced interference patterns, it is sufficient to 
consider the periodicity of the lattice, and to identify the probability 
distribution for the occupation of a given quasimomentum state. We have 
demonstrated the above statement by reproducing a number of features of 
the interference patterns with remarkable accuracy using an approximation 
where we neglect those density-density correlations. In particular, the 
average visibility decreases as $N^{-1/2}$ with increasing number of 
condensates $N$. Our results help explain the behaviour observed in recent 
experiments on the interference of atomic clouds released from optical 
lattices.

The author would like to thank R. Diener and T.-L. Ho for useful 
discussions. This work was supported by the National Science Foundation 
through Grant Nos. DMR-0071630 and DMR-0109255 and by NASA through Grant 
Nos. NAG8-1441 and NAG8-1765.

\vspace{2cm}

\noindent {\bf Figures}

Fig. 1: Average visibility $V$ as a function of number of condensates $N$ 
on a logarithmic scale. Circles and stars represent the exact calculation 
for the Thomas-Fermi and the homogeneous case, respectively. X's are 
generated by summing $N$ random numbers of unit modulus at each value of 
quasimomentum. +'s are generated using the asymptotic two-dimensional 
random walk probability distribution at each value of quasimomentum.

Fig. 2: Probability density $P(V)$ as a function of visibility $V$ for 
$N=11$ (a) and $N=51$ (b). X's and +'s are obtained from the exact 
calculation and the calculation where density-density correlations are 
neglected, respectively. The solid line is the asymptotic two-dimensional 
random walk probability density corresponding to a mean distance 
$\sqrt{\pi/11}=0.53$ (a) and $\sqrt{\pi/51}=0.25$ (b).


\begin{thebibliography}{99}
\bibitem{NumberOfAtoms} As we shall explain below, the atoms in each 
quasimomentum state, whose number is of the same order as the number of 
atoms per lattice site, are distributed among the different interference 
fringes. Therefore, in order to be able to see the periodic density 
distribution, one must have a large number of atoms per quasimomentum 
state, so that number fluctuations can be neglected.
\end{thebibliography}
\end{document}